\documentclass[apj]{emulateapj}
\def\b{b}
\newcommand{\pdeg}{.\negthickspace^\circ}

\newcommand{\gcc}{\mathrm{g\,cm^{-3}}}

\newcommand{\lfill}{\leavevmode%
\leaders\hrule depth-2.1pt height 2.3pt\hfill\kern0pt
}
\usepackage[dvipdfmx]{color}
\usepackage{bm}
\usepackage{aas_macros} 
\usepackage{float}
\usepackage{amsmath,amssymb}
\usepackage{fancybox}
\usepackage{url}
\usepackage{booktabs}
\usepackage{threeparttable}

\slugcomment{Accepted to ApJ}
\shortauthors{Masuda}
\shorttitle{Spin--Orbit Angles from Stellar Gravity Darkening}
\begin{document}
\title{SPIN--ORBIT ANGLES OF KEPLER-13A${\rm \b}$ AND HAT-P-7${\rm \b}$\\FROM GRAVITY-DARKENED TRANSIT LIGHT CURVES}
\author{
Kento \textsc{Masuda}\altaffilmark{\dag}
} 
\affil{
Department of Physics, The University of Tokyo, Tokyo 113-0033, Japan
}
\email{$\dag$ masuda@utap.phys.s.u-tokyo.ac.jp}
\begin{abstract}
Analysis of the transit light curve deformed by the stellar gravity darkening 
allows us to photometrically measure both components 
of the spin--orbit angle $\psi$, its sky projection $\lambda$
and inclination of the stellar spin axis $i_\star$.
In this paper, we apply the method to two transiting hot Jupiter systems monitored
with the {\it Kepler} spacecraft, Kepler-13A and HAT-P-7.
For Kepler-13A, we find $i_\star=81^\circ\pm5^\circ$ and $\psi=60^\circ\pm2^\circ$
adopting the spectroscopic constraint $\lambda=58\pdeg6\pm2\pdeg0$ by Johnson et al. (2014). 
In our solution, the discrepancy between the above $\lambda$ 
and that previously reported by Barnes et al. (2011) is solved
by fitting both of the two parameters in the quadratic limb-darkening law.
We also report the temporal variation in the orbital inclination of Kepler-13Ab,
$\mathrm{d} |\cos i_{\rm orb}|/\mathrm{d}t=(-7.0\pm0.4)\times10^{-6}\,\mathrm{day}^{-1}$, 
providing further evidence for the spin--orbit precession in this system.
By fitting the precession model to the time series of $i_{\rm orb}$, $\lambda$, and $i_\star$
obtained with the gravity-darkened model,
we constrain the stellar quadrupole moment $J_2=(6.1\pm0.3)\times10^{-5}$ for our new solution,
which is several times smaller than $J_2=(1.66\pm0.08)\times10^{-4}$ obtained for the previous one.
We show that the difference can be observable in the future evolution of $\lambda$,
thus providing a possibility to test our solution with follow-up observations.
The second target, HAT-P-7, is the first F-dwarf star analyzed with the gravity-darkening method.
Our analysis points to a nearly pole-on configuration with $\psi=101^\circ\pm2^\circ$ or $87^\circ\pm2^\circ$
and the gravity-darkening exponent $\beta$ consistent with $0.25$.
Such an observational constraint on $\beta$ can be useful for testing the theory of gravity darkening.
\end{abstract}

\section{Introduction}\label{sec:intro}
Spin--orbit angle or the stellar obliquity, $\psi$, the angle between the stellar spin axis and the orbital axis of its planet,
serves as a unique probe of the dynamical history of planetary systems.
Especially, its connection with the hot-Jupiter migration has been extensively studied \citep[e.g.,][]{2009ApJ...696.1230F},
but the relationship between the observed samples and the migration process is not straightforward
for various reasons.
First of all, the initial distribution of the spin--orbit angles is not known.
Some studies do suggest that the protoplanetary disk may have already been misaligned with the stellar equator
due to the chaotic gas accretion \citep[e.g.,][]{2010MNRAS.401.1505B, 2014arXiv1409.5148F}
or the magnetic star--planet interaction \citep[e.g.,][]{2011MNRAS.412.2790L}.
In these cases, the spin--orbit misalignment is primordial, rather than due to the migration.
Even after the disk dissipation or the completion of migration, spin--orbit angle can evolve
due to the gravitational perturbation from the companion \citep[e.g.,][]{2014Sci...345.1317S, 2014ApJ...794..131L}.
As suggested by the observed correlation between the spin--orbit misalignment and stellar effective temperature \citep{2010ApJ...718L.145W, 2012ApJ...757...18A},
spin--orbit angle may also be affected by the tidal star--planet interaction \citep[e.g.,][]{2014ApJ...784...66X}, 
whose mechanism is not well understood.
To partially resolve these issues, it is beneficial to measure spin--orbit angles for systems
with various host-star and orbital properties.
For instance, planets on distant orbits or around hot/young stars are valuable targets
because we expect that tides have not significantly affected the primordial spin--orbit configuration.

This paper focuses on a relatively new method for the spin--orbit angle determination
in transiting systems,
which utilizes the gravity darkening of the host star owing to its rapid rotation \citep{2009ApJ...705..683B}.
Stellar rotation makes the effective surface gravity at the stellar equator smaller than that at the pole
by a fractional order of $\gamma \equiv \Omega_\star^2 R_\star^3/2GM_\star \sim (P_{\rm br}/P_{\rm rot})^2$,
where $\Omega_\star$, $R_\star$, $M_\star$, $P_{\rm br}$, and $P_{\rm rot}$
are angular rotation frequency, radius, mass, break-up rotation period, and rotation period of the star,
respectively.
According to von Zeipel's theorem \citep{1924MNRAS..84..665V},
this results in the inhomogeneity of the stellar surface brightness through the relation
$T_{\rm eff} \propto g_{\rm eff}^{\beta}$.
Here, $T_{\rm eff}$ and $g_{\rm eff}$ are 
the effective temperature and surface gravity at each point on the stellar surface,
and gravity-darkening exponent $\beta$ characterizes the strength of the gravity darkening,
which is theoretically $0.25$ for a barotropic star with a radiative envelope.
When a planet transits a star with such an inhomogeneous and generally non-axisymmetric brightness distribution,
an anomaly of $\mathcal{O}(\gamma \delta)$ appears in the light curve,
where $\delta$ is the transit depth.
Since the shape of the anomaly depends on the position of the stellar pole
relative to the planetary orbit, the obliquity of the stellar spin $\psi$ can be estimated
with the light-curve model taking into account the effect of gravity darkening.

Indeed, this ``gravity-darkening method" has many unique aspects.
So far, it is the only known method that simultaneously constrains {\it both} 
components of $\psi$, the 
sky-projected spin--orbit angle $\lambda$ and stellar inclination $i_\star$
(c.f., Equation \ref{eq:psi} and Figure \ref{fig:angles}).
Moreover, obliquity analysis is possible essentially with the photometric data alone,
and its application is not necessarily limited to short-period planets, as far as the transit is observed
with sufficient signal-to-noise ratio \citep{2013ApJ...776L..35Z}.
It is also interesting to note that the method is (only) applicable to fast-rotating (i.e., young or hot) stars,
for which anomalies of larger amplitudes result.
Since rapid rotators are not suitable for the precise spectroscopic velocimetry because of their
broad spectral lines, this method is complementary to the conventional 
spin--orbit angle measurement using the Rossiter-McLaughlin (RM) effect.
All these properties make the method suitable for sampling stars for which tidal effect
is not so significant that the primordial information is expected to be well preserved
in the current spin--orbit configuration.

Although the gravity-darkening method is valuable in many aspects, 
the procedure for obtaining $\psi$ may not be fully established. 
In a representative example of its application, Kepler-13A,
the constraint from the gravity-darkening method \citep[][hereafter B11]{2011ApJS..197...10B} 
is known to be in disagreement with the later spectroscopic measurement of $\lambda$ 
with the Doppler tomography \citep{2014ApJ...790...30J}.
In addition, inconsistent results arise even within the gravity-darkening analyses,
depending on the choice of the limb-darkening coefficients
or $\beta$ \citep{2013ApJ...776L..35Z, 2014ApJ...786..131A}.
For these reasons, it is worth revisiting the reliability and limitation
of this method more carefully,
in order for this unique method to be applied to more systems in future and provide credible results.

In this paper, we reanalyze a well-known example of the gravity-darkened transit of Kepler-13Ab, 
with more data than used in the previous analysis by B11.
We investigate the systematic effects in the spin--orbit angle determination,
and propose a joint solution that may solve the discrepancy with the Doppler tomography measurement
(Section \ref{sec:koi13}).
We will also see that the spin--orbit precession in this system can be used to 
test the validity of our solution, as well as to determine the stellar quadrupole moment $J_2$
(Section \ref{sec:precession}).

In addition, we apply the gravity-darkening method for the first time to an F-type dwarf star,
HAT-P-7, where the anomaly in the transit light curve has been reported in several studies 
\citep[e.g.,][]{2013ApJ...772...51E, 2013ApJ...774L..19V, 2014arXiv1407.2245E, 2014PASJ...66...94B}.
While the RM measurements \citep{2009ApJ...703L..99W, 2009PASJ...61L..35N, 2012ApJ...757...18A} 
have established that $\lambda>90^\circ$, suggesting a retrograde orbit,
the following asteroseismic inferences \citep{2014PASJ...66...94B, 2014A&A...570A..54L}
have revealed that a pole-on orbit is actually favored.
In Section \ref{sec:hatp7}, we show that a similar conclusion is also obtained from the gravity-darkening method
and discuss the consistency of our result with other constraints on the host-star properties.

\section{Method}\label{sec:method}
\subsection{Model}\label{ssec:method_model}
We basically follow \citet{2009ApJ...705..683B} in modeling the gravity-darkened transit light curve.
The model includes the following 14 parameters, 
which are listed as ``fitting parameters" in Tables \ref{tab:koi13_gdfit} and \ref{tab:hatp7}:
\begin{enumerate}
\item mean stellar density, $\rho_{\star} = 3M_\star/4\pi R_\star^3$, 
	which corresponds to the semi-major axis scaled by the stellar equatorial radius, $a/R_\star$\footnote{In this paper, $R_\star$ denotes the equatorial radius of the star.}
\item limb-darkening coefficient for the quadratic law, $c_1=u_1+u_2$,
\item limb-darkening coefficient for the quadratic law, $c_2=u_1-u_2$,
\item time of the inferior conjunction, $t_c$,
\item orbital period, $P$,
\item cosine of orbital inclination, $\cos i_{\rm orb}$,
\item planetary radius normalized to the stellar equatorial radius, $R_{\rm p}/R_\star$
\item normalization of the out-of-transit flux, $F_0$
\item stellar mass, $M_\star$,
\item stellar rotation frequency, $f_{\rm rot}$
\item stellar effective temperature at the pole, $T_{\star, \rm pole}$
\item gravity-darkening exponent, $\beta$,
\item stellar inclination, $i_\star$
\item sky-projected spin--orbit angle, $\lambda$.
\end{enumerate}

The first eight parameters are common with the light-curve model without gravity darkening.
We assume circular orbits for the two targets because the orbital eccentricities are
constrained to be very small, if any, from the occultation light curves 
\citep{2014ApJ...788...92S, 2014PASJ...66...94B}.

In the gravity-darkened model by \citet{2009ApJ...705..683B}, the shape of the star is approximated by the spheroid with the oblateness
$\gamma = \Omega_\star^2 R_\star^3/2GM_\star = 3\pi f_{\rm rot}^2 / 2G\rho_\star$.
The surface brightness at each point is modeled as the blackbody emission of the temperature
$T_\star = T_{\star, \rm pole} \left( g_{\rm eff}/g_{\rm eff, pole} \right)^\beta,$
where $g_{\rm eff}/g_{\rm eff, pole}$ is the effective surface gravity normalized by its value at the stellar pole.
The surface gravity at point $\bm{r}$ on the stellar surface is calculated by 
$\bm{g}_{\rm eff} = - GM_\star r^{-2} \bm{\hat r} + 4\pi^2 f_{\rm rot}^2 r_\perp \bm{\hat r_\perp}.
\label{eq:g_vector}$
Here $r$ and $\bm{\hat r}$ are the norm and unit vector of the radius vector $\bm{r}$, respectively.
Similarly, $r_\perp$ and $\bm{\hat r_\perp}$ are those of $\bm{r_\perp}$,
the projection of $\bm{r}$ onto the stellar equatorial plane.
The Planck function $B_\lambda(T_\star)$ at each point is convolved with the 
``high-resolution" {\it Kepler} response function\footnote{\url{http://keplergo.arc.nasa.gov/CalibrationResponse.shtml}}
using the table of the wavelength- and temperature-dependent factor calculated prior to the fitting.
The convolved flux is then multiplied by the limb-darkening function 
\begin{equation}
	I(\mu) = 1 - u_1 (1 - \mu) - u_2 (1 - \mu)^2,
\end{equation}
with $\mu$ being the cosine of the angle between $-\bm{g}_{\rm eff}$ and our line of sight,\footnote{
Although this vector $-\bm{g}_{\rm eff}$ is not exactly parallel to the surface normal of the spheroid we assume, the difference is $\mathcal{O}(\gamma^2)$ and thus negligible.}
and integrated over the visible surface of the star to give the total flux.
We fix $T_{\star, \rm pole}$ at the observed effective temperature
assuming that the difference between $T_{\star, \rm pole}$ and the disk-integrated effective temperature is small.
Note that the gravity-darkened transit light curve gives $\rho_\star$ alone
and can not constrain $M_\star$ and $R_\star$ separately,
as is the case for the transit without gravity darkening.

\begin{figure}
	\centering
	\includegraphics[width=8cm,clip]{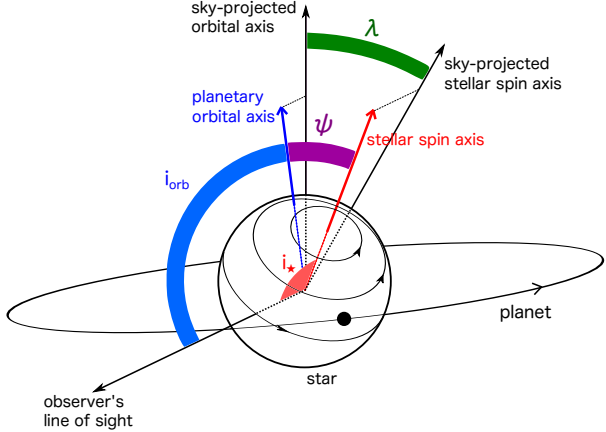}
	\caption{Definitions of $i_{\rm orb}$, $i_\star$, $\lambda$, and $\psi$ in this paper.
	The orbital inclination, $i_{\rm orb}$, is the angle between the planetary orbital axis
	(blue arrow) and the observer's line of sight. In a transiting system, $i_{\rm orb}$ is usually very close to $\pi/2$ 
	and hence the orbital axis almost coincides with its projection onto the plane of the sky.
	Inclination of the stellar spin axis, $i_\star$, is similarly defined as 
	the angle between the stellar spin axis (red arrow) and the line of sight.
	The angle between the two axes (red and blue ones), $\psi$, is the spin--orbit angle or the stellar obliquity.
	Its sky projection, $\lambda$, denotes the angle between the sky projections of the
	same two axes.}
	\label{fig:angles}
\end{figure}

The configuration of the planetary orbit and stellar spin is specified by three angles,
$i_{\rm orb}$, $i_\star$, and $\lambda$,
which are defined in Figure \ref{fig:angles}
\citep[see also figure 1 of][]{2014PASJ...66...94B}.
The orbital and stellar inclinations, $i_{\rm orb}$ and $i_\star$, 
are measured from the line of sight and defined to be in the range $[0, \pi]$.
The sky-projected spin--orbit angle, $\lambda$, is the angle between the sky-projected stellar spin 
and planetary orbital axes.
It is measured from the former to the latter counterclockwise in the sky plane, and is in the range $[0, 2\pi]$.
With these definitions, the true spin--orbit angle, or the stellar obliquity, $\psi$, 
is given by equation 1 of \citet{2014PASJ...66...94B}:
\begin{equation}
	\label{eq:psi}
	\cos \psi = \cos i_\star \cos i_{\rm orb} + \sin i_\star \sin i_{\rm orb} \cos \lambda.
\end{equation}
Throughout the paper, we restrict $i_\star$ to be in the range $[0, \pi/2]$
making use of the intrinsic symmetry with respect to the sky plane.
We do not lose any physical information of the system with this choice 
because any of the relative star--planet configurations with $i_\star$ in $[\pi/2, \pi]$
is the same as one of those with $i_\star$ in $[0, \pi/2]$.
In other words, the configurations $(i_\star, i_{\rm orb}, \lambda)$ and $(\pi-i_\star, \pi-i_{\rm orb}, -\lambda)$ are equivalent.
This transformation corresponds to looking at the system from the other side of the plane of the sky.

In the following, we also adopt the constraint on the stellar line-of-sight 
rotational velocity $v\sin i_\star$ from spectroscopy,
which is related to the above model parameters by
\begin{equation}
	\label{eq:vsini}
	v\sin i_\star 
	= 2\pi f_{\rm rot} \left( \frac{3M_\star}{4\pi \rho_\star} \right)^{1/3}  \sin i_\star.
\end{equation}
This, in principle, allows us to break the degeneracy between $M_\star$ and $R_\star$, enabling 
the determination of the absolute dimension of the system.
Nevertheless, the constraint on $M_\star$ is usually weak as discussed in B11, and so we fix $M_\star$ 
at the observed value.

\subsection{Data Processing}\label{ssec:method_data}
We detrend and normalize the transit light curves of each target
along with the consistent determination of the transit times and transit parameters.
We first normalize the light curve of each quarter using its median, 
and then iterate the following two steps until the resulting transit times $t_c$
and transit parameters converge (typically 10--20 times):
\begin{enumerate}
\item Light curve around each transit ($\pm 0.2\,\mathrm{days}$ for Kepler-13A and $\pm 0.15\,\mathrm{days}$ for HAT-P-7)
is modeled as the product of a quadratic polynomial\footnote{
Use of the quadratic polynomial helps the better removal of flux variation not due to the transit,
i.e., planetary light, ellipsoidal variation, and Doppler beaming.}
$a_0+a_1 (t-t_c) + a_2 (t-t_c)^2$ ($t$: time) 
and the analytic transit light-curve model by \citet{2002ApJ...580L.171M}.
We use the Levenberg-Markwardt (LM) method \citep{2009ASPC..411..251M}
to fit $a_0$, $a_1$, $a_2$, and $t_c$ iteratively removing $5\sigma$ outliers, while the other parameters are fixed.
The filtered data are then divided by the best-fit polynomial to give a normalized and detrended transit light curve.
We discard the transits with data gaps of more than $50\%$.
\item Using the set of $t_c$ obtained in the first step, we calculate the mean orbital period $P$ and 
transit epoch $t_0$ by linear fit
and use them to phase-fold the normalized and detrended transits.
The phase-folded light curve is averaged into one-minute bin and then fitted 
with the \citet{2002ApJ...580L.171M} model using an LM algorithm.
We fit $c_1$, $c_2$, $\rho_\star$, $\cos i_{\rm orb}$, $R_{\rm p}/R_\star$, and $F_0$,
whose best-fit values are used in the step 1 of the next iteration.
In this step, the orbital period $P$ is fixed to be the value obtained from the linear fit and the central time
of the phase-folded transit is fixed to be zero.
\end{enumerate}

In the following analysis, we use the one-minute binned, phase-folded light curve 
obtained in the second step of the final iteration.
For each bin, the flux value is given by its mean and the error is
estimated as the standard deviation within the bin divided by the 
square root of the number of data points.

\subsection{Fitting Procedure}\label{ssec:method_fit}
In fitting the observed light curves, the likelihood $\mathcal{L}$ of the model 
is computed by $\mathcal{L}\propto\exp(-\chi^2/2)$, where
\begin{equation}
	\label{eq:chi2}
	\chi^2 = \sum_{i} \left( \frac{f_i - f_{\mathrm{model}, i}}{\sigma_i} \right)^2
	+ \sum_{j} \left( \frac{p_j - p_{\mathrm{model}, j}}{\delta p_j} \right)^2.
\end{equation}
In the first term, $f_i$, $f_{\mathrm{model},i}$, and $\sigma_i$ are the observed value, 
modeled value, and error of the $i$th flux data.
The second term is introduced to take into account the constraints from other observations on some 
(functions) of the model parameters ${p_j}$.
In the following analysis, $p$ is read to be $v\sin i_\star$ and, in some cases, $\lambda$.\footnote{
Only in Section \ref{ssec:tpars}, $\rho_\star$, $c_1$, $c_2$, $R_{\rm p}/R_\star$, and $f_{\rm rot}$
are also included.}
For each $p_j$, we assume a Gaussian constraint of the form $p_j \pm \delta p_j$ 
and the value obtained from the model is denoted by $p_{\mathrm{model}, j}$.

The maximum likelihood solution is found by
minimizing Equation (\ref{eq:chi2}) with the LM method using the {\tt cmpfit} package \citep{2009ASPC..411..251M}.
Since the complex dependence of $\chi^2$ on $i_\star$ and $\lambda$ is expected, 
we repeat the fitting procedure from the initial $i_\star$ in $[0, 90^\circ]$
and $\lambda$ in $[-180^{\circ}, 180^{\circ}]$ at $10^{\circ}$ intervals.
Initial values of the other parameters are chosen close to the best-fit values 
obtained from the model without gravity darkening.
We also try both positive and negative $\cos i_{\rm orb}$ as an initial value 
to search the whole domain of $i_{\rm orb}$, which is now $[0^\circ, 180^\circ]$.

\section{Transit Analysis of Kepler-13A\b}\label{sec:koi13}
In this section, we report the analysis of the gravity-darkened transit of Kepler-13Ab.
We first analyze the whole available short-cadence (SC) data from Q2, 3, and 7--17 using the same
stellar parameters as in B11
to test the validity of our method (Section \ref{ssec:koi13_b11}).
Motivated by the recently reported disagreement with $\lambda$ from the Doppler tomography,
we also investigate the possible systematics in the spin--orbit determination
arising from the choice of stellar parameters.
We show that the discrepancy can be absorbed by adjusting the value of $c_2$
and present a joint solution that is compatible with all of the observations made so far.

\subsection{Reproducing the Results by B11}\label{ssec:koi13_b11}
In this subsection, we analyze the short-cadence (SC), 
Pre-search Data Conditioned Simple Aperture Photometry (PDCSAP) fluxes from Q2, 3, and 7--17.
Given the clear transit duration variation (TDV) reported by \citet{2012MNRAS.421L.122S} and \citet{2014MNRAS.437.1045S},
we separately analyze the transits from each quarter, rather than folding all the available data.
Since we do not detect significant temporal variations in the parameters other than $\cos i_{\rm orb}$ (see
Section \ref{sec:precession}),
we report the mean and standard deviation of the best-fit values from the above 13 quarters for each parameter. 

\begin{figure*}
	\centering
	\includegraphics[width=12cm, bb=50 50 280 302]{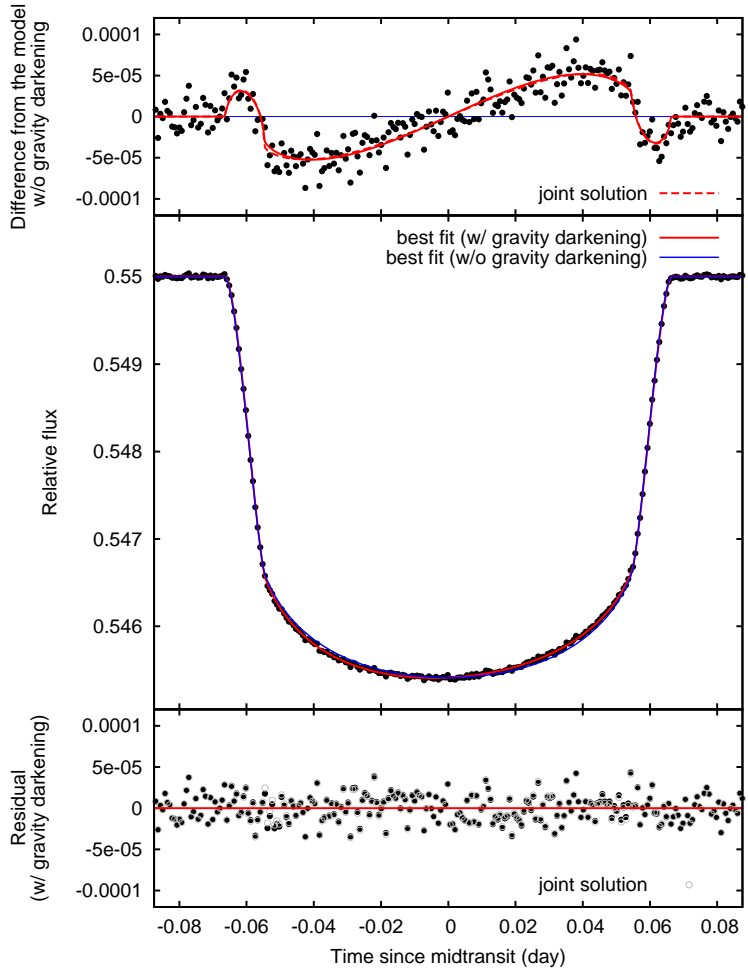}
	\caption{Fitting the gravity-darkened model to the Q2 transit of Kepler-13Ab.
	(Middle) Black dots are the phase-folded and binned fluxes from Q2. 
	The thick red line shows the best-fit gravity-darkened model,
	while the thin blue line is the best-fit model without gravity darkening.
	(Bottom) Black dots are the residual of the best-fit gravity-darkened model. 
	Gray open circles are those for the joint solution,
	where $c_2$ is fitted with the constraint $\lambda=58\pdeg6 \pm 2\pdeg0$
	from the Doppler tomography.
	(Top) Black dots are the residuals of the best-fit model without gravity darkening.
	Thick red line is the difference between the best-fit model with gravity darkening and that without gravity darkening.
	Dashed red line shows the same result for the joint solution.
	The difference between the two gravity-darkened solutions
	is only barely visible just after the ingress and before the egress.}
	\label{fig:koi13_best_q2}
\end{figure*}

First, we use the same stellar parameters as in B11 and obtain the results in the second column of Table \ref{tab:koi13_gdfit}.
Namely, we subtract a constant value $F_{\rm c}=0.45$ from the normalized flux to 
remove the flux contamination from the companion star,
and impose the constraint $v\sin i_\star = 65\pm10\,\mathrm{km\,s^{-1}}$ based on \citet{2011ApJ...736L...4S}.
We fix $M_\star = 1.83M_\odot$ and $T_{\star, \rm pole} =8848\,\mathrm{K}$ from \citet{2011ApJ...736...19B},
and $c_2=0$.
In Figure \ref{fig:koi13_best_q2}, the best-fit model is overplotted with the data for Q2,
which is to be compared with figure 2 of B11.

Basically, we find a very good agreement with the result by B11 using about $12$ times more data.
Although the values of $\cos i_{\rm orb}$, $i_\star$, and $\lambda$ we report here appear
different from those in B11,
that is simply because we choose $i_\star$ to be in the range $[0, \pi/2]$.
This is physically the same configuration as theirs 
and corresponds to the top-left situation in figure 3 of B11.
That is, \emph{$\lambda$ in our solutions with $\cos i_{\rm orb}<0$
should be read as $-\lambda$ in the conventional definition, 
because $\lambda$ is usually defined for the orbit with $\cos i_{\rm orb}>0$}
(see also the discussion after Equation \ref{eq:psi}).

In addition to the solution in Table \ref{tab:koi13_gdfit}, 
we also find a retrograde solution with $\lambda>90^{\circ}$ as noted in B11.
Here we do not discuss this solution, however, because the Doppler tomography observation
has already excluded the retrograde orbit with high significance \citep{2014ApJ...790...30J}.
\subsection{Systematics due to Stellar Parameters}\label{ssec:koi13_systematic}
Although we find consistent values of $\lambda$ and $i_\star$ as obtained by B11,
those of $\lambda$ significantly differ from $\lambda=58\pdeg6 \pm 2\pdeg0$,
the value obtained from the Doppler tomography \citep{2014ApJ...790...30J}.
Motivated by this discrepancy, 
we investigate the possible origins of systematics in the spin--orbit angle determination
with gravity darkening in this subsection.

First, we examine the systematics due to the choice of 
$M_\star$, $v \sin i_\star$, $T_{\star, \rm pole}$, and $F_{\rm c}$,
which are the stellar properties not derived from the light curve modeling.\footnote{
We do not examine the dependence on $\beta$ here
because B11 have already shown that a different choice of $\beta=0.19$,
suggested by the interferometric observation of Altair \citep{2007Sci...317..342M}, does not change the result significantly.}
We perform the same analysis as in Section \ref{ssec:koi13_b11}, but adopting the following parameters from
the most recent photometric and spectroscopic study by \citet[][hereafter S14]{2014ApJ...788...92S}:
$v\sin i_\star = 78\pm15\,\mathrm{km\,s^{-1}}$,
$M_\star=1.72M_\odot$, 
$T_{\star,\rm pole}=7650\,\mathrm{K}$, and
$F_{\rm c}=0.47726$.
The corresponding results are shown in the third column of Table \ref{tab:koi13_gdfit}. 
We find that $i_\star$ and $\lambda$ can differ by as large as $10^{\circ}$ 
due to the choice of the above parameters, 
but the difference is not so large as to explain the disagreement with the Doppler tomography.
The main difference from the B11 case with this new set of parameters 
is the different constraint on $f_{\rm rot} \sin i_\star$,
which is proportional to the combination $(\rho_\star/M_\star)^{1/3} v \sin i_\star$ (c.f., Equation \ref{eq:vsini}).
With smaller $M_\star$ and larger $v\sin i_\star$, 
the stellar rotation rate slightly higher than the B11 case is favored.
We find that the difference in $T_{\star, \rm pole}$ is less important
compared to the above effect.
We also find that larger $F_{\rm c}$ yields larger $R_{\rm p}/R_\star$, 
which makes the impact parameter or $|\cos i_{\rm orb}|$ smaller to give the same
ingress/egress duration.

\begin{figure*}[htbp]
	\centering
	\includegraphics[width=10.5cm,clip]{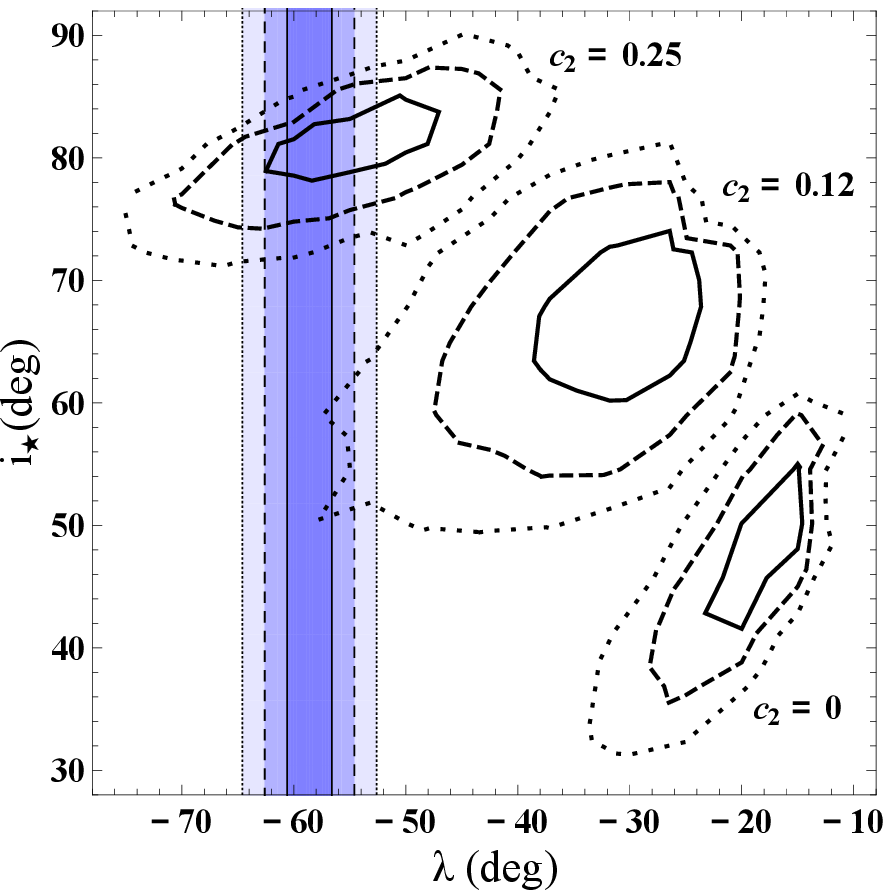}
	\caption{Constraints on $(\lambda, i_\star)$ from the gravity-darkened transit of Kepler-13Ab for 
	the different choices of $c_2$.
	In this illustration, data from Q2 are used and stellar parameters from B11 are adopted.
	The solid, dashed, and dotted contours respectively show $1\sigma$, $2\sigma$, and $3\sigma$ confidence regions for $(\lambda, i_\star)$
	obtained from $200000$ Markov Chain Monte Carlo (MCMC) samples for three fixed values of $c_2$ ($0$, $0.12$, and $0.25$).
	The shaded areas bounded by the vertical solid, dashed, and dotted lines 
	respectively denote $1\sigma$, $2\sigma$, and $3\sigma$ confidence regions for $\lambda$
	obtained from the Doppler tomography \citep{2014ApJ...790...30J}.
	The sign of $\lambda$ is opposite to their quoted value because we are now dealing with the solution
	with $\cos i_{\rm orb}<0$ (i.e., $\pi/2<i_{\rm orb}<\pi$); 
	see also the discussion in the third paragraph of Section \ref{ssec:koi13_b11}.}
	\label{fig:is_lambda_correlation}
\end{figure*}

Next, we allow $c_2=u_1-u_2$ to be free, 
and find that the resulting spin--orbit angle is very sensitive to this parameter.
When $c_2$ is floated, 
the constraints on $i_\star$ and $\lambda$ become much weaker than the $c_2=0$ case,
as shown in the fourth and fifth columns of Table \ref{tab:koi13_gdfit}.
The strong dependence on $c_2$ is illustrated in Figure \ref{fig:is_lambda_correlation},
which shows that $\lambda$ and $i_\star$ vary by several tens of degrees depending on $c_2$.
In fact, the result indicates that the gravity-darkened light curve is actually compatible 
with the Doppler tomography solution 
if we choose $c_2\sim0.25$; such a solution will be discussed in Section \ref{ssec:koi13_joint}.

\subsection{Joint Solution}\label{ssec:koi13_joint}
In Section \ref{ssec:koi13_systematic}, we found that the gravity-darkened light curve is
compatible with the value of $\lambda$ estimated from the Doppler tomography
if $c_2 \sim 0.25$.
Thus we repeat the analysis treating $c_2$ as a free parameter 
for both stellar parameters by B11 and S14, but this time 
imposing additional constraint $\lambda=58\pdeg6\pm2\pdeg0$ from the Doppler tomography.
The results are summarized in the last two columns in Table \ref{tab:koi13_gdfit}.
The resulting value of $i_\star=81^\circ\pm5^\circ$ indicates that the star is close to equator-on,
and $\psi=60^\circ\pm2^\circ$ is slightly larger than the previous estimate.
In terms of $\chi^2_{\rm min}$, 
these solutions equally well reproduce the transit anomaly as the solutions discussed so far,
and still they are consistent with the Doppler tomography result.
Moreover, we obtain a slightly longer $P_{\rm rot}$, which better agrees with 
$P_{\rm rot}=25.43\pm0.05\,\mathrm{h}$ 
estimated by \citet{2012MNRAS.421L.122S} and \citet{2014MNRAS.437.1045S} 
than the solution with the gravity darkening alone. 
For these reasons, the joint solution is most favored from the current observations.

We note, however, that the likelihood for the joint solution is not so high as
to statistically justify the introduction of the additional free parameter $c_2$. 
Furthermore, the plausibility of the value of $c_2$ in our joint solution is theoretically unclear.
We obtain the theoretical values of $c_{1, \rm th}\simeq0.6$ and $c_{2, \rm th}\simeq0.0$ 
from the table of \citet{2010A&A...510A..21S} if we adopt the effective temperature and surface gravity by S14.
Hence the value of $c_2$ from our joint solution is discrepant from $c_{2, \rm th}$;
they could even have opposite signs depending on the stellar parameters.
Nevertheless, it is also true that theoretical values often disagree with the observed ones \citep[e.g.,][]{2008MNRAS.386.1644S};
in fact, $c_1$ in the light-curve solution with $c_2=0$ is also different from $c_{1, \rm th}$.
Therefore, we do not consider the possible deviations from the theoretical values crucial,
and regard it as an open question.\footnote{For reference, we find $c_2=0.1\mathchar`-0.2$ 
if we adopt the model without gravity darkening \citep{2002ApJ...580L.171M}, 
which suggests that the choice of $c_2=0$ is not indispensable.}
An alternative approach to independently 
assess the validity of our solution is discussed in the next section.

\begin{deluxetable*}{lcccccc}
	\tabletypesize{\small}
	\tablewidth{0pt}
	\tablecolumns{7}
	\tablecaption{Results for the transit of Kepler-13A${\rm \b}$}
	\tablehead{
	& \multicolumn{2}{c}{light-curve solution ($c_2=0$)}
	& \multicolumn{2}{c}{light-curve solution ($c_2$ fitted)}
	& \multicolumn{2}{c}{joint solution ($c_2$ fitted)}\vspace{0.1cm}\\
	\colhead{Ref. for $F_{\rm c}$, $v\sin i_\star$, $M_\star$, $T_{\star,\rm pole}$}  
	& \colhead{B11} & \colhead{S14} 
	& \colhead{B11} & \colhead{S14}
	& \colhead{B11} & \colhead{S14}
	}
	\startdata
	({\it Assumed Flux Contamination})\\
	$F_{\rm c}$ 	&	$0.45$	&	$0.47726$	&	$0.45$	&	$0.47726$ &	$0.45$	&	$0.47726$\\
	\vspace{-0.2cm}\\
	\multicolumn{1}{l}{({\it Constraints})}\\
	$v\sin i_\star$ ($\mathrm{km\,s^{-1}}$) & $65\pm10$ & $78\pm15$ & $65\pm10$ & $78\pm15$ & $65\pm10$ & $78\pm15$\\
	$\lambda$ (deg)	 & $\cdots$ & $\cdots$ & $\cdots$ & $\cdots$ & $58.6\pm2.0$\tablenotemark{*} & $58.6\pm2.0$\tablenotemark{*}\\
	\vspace{-0.2cm}\\
	\multicolumn{1}{l}{({\it Fitting Parameters})}\\
	$M_\star$ ($M_\odot$) 	 & $1.83$ (fixed) & $1.72$ (fixed) & $1.83$ (fixed) & $1.72$ (fixed) & $1.83$ (fixed) & $1.72$ (fixed)\\
	$T_{\star, \rm pole}$ (K) 	& $8848$ (fixed) & $7650$ (fixed) & $8848$ (fixed) & $7650$ (fixed) & $8848$ (fixed) & $7650$ (fixed)\\
	$\rho_\star$ ($\mathrm{g\,cm^{-3}}$)	 & $0.533\pm0.005$ & $0.550\pm0.005$ & $0.530\pm0.005$ & $0.547\pm0.006$ & $0.525\pm0.005$ & $0.538\pm0.006$\\
	$c_1$	  & $0.496\pm0.008$ & $0.493\pm0.008$ & $0.50\pm0.04$ & $0.51\pm0.02$ & $0.523\pm0.005$ & $0.528\pm0.006$\\
	$c_2$	  & $0$ (fixed) & $0$ (fixed) & $0.02\pm0.28$ & $0.12\pm0.13$ & $0.20\pm0.02$ & $0.26\pm0.04$\\
	$t_c$ ($10^{-5}$day)\tablenotemark{**} & $-3\pm1$ & $-3\pm1$ & $-8\pm7$ &$-10\pm7$ & $-9\pm4$ & $-11\pm5$\\
	$P$ (day)	 & \multicolumn{6}{c}{\lfill\ $1.763587\pm0.000002$\ \lfill}\\
	$\cos i_{\rm orb}$	 & $-0.066\pm0.004$ & $-0.057\pm0.004$ & $-0.066\pm0.003$ & $-0.055\pm0.004$ & $-0.064\pm0.004$ & $-0.054\pm0.004$\\
	$R_{\rm p}/R_\star$	  & $0.0845\pm0.0002$ & $0.0864\pm0.0002$ & $0.0845\pm0.0003$ & $0.0865\pm0.0002$ & $0.0846\pm0.0002$ & $0.0864\pm0.0003$\\
	$F_0$	 &\multicolumn{6}{c}{\lfill\ $0.550000\pm0.000002$ (B11) / $0.522740\pm0.000002$ (S14)\ \lfill}\\
	$f_{\rm rot}$ ($\mu\mathrm{Hz}$)	  & $12.9\pm0.4$ & $14.5\pm0.6$ & $12.8\pm4.2$ & $12.9\pm1.8$ & $10.2\pm0.6$ & $11.6\pm1.0$\\
	$i_\star$ (deg) 	  & $47\pm3$ & $56\pm3$ & $60\pm20$ & $71\pm16$ & $73\pm5$ & $81\pm5$\\
	$\lambda$ (deg)	  & $-20.3\pm1.3$ & $-13.9\pm1.3$ & $-33\pm13$ & $-30\pm12$ & $-58.4\pm2.0$\tablenotemark{***} & $-58.5\pm2.0$\tablenotemark{***}\\
	$\beta$	 & $0.25$ (fixed) & $0.25$ (fixed) & $0.25$ (fixed)  & $0.25$ (fixed)  & $0.25$ (fixed) & $0.25$ (fixed)\\
	\vspace{-0.2cm}\\
	\multicolumn{1}{l}{({\it Derived Parameters})}\\
	$P_{\rm rot}$ (hr)	& $21.5\pm0.7$	& $19.1\pm0.8$	& $23\pm5$ 		& $22\pm3$ 	& $27\pm2$	& $24\pm2$\\
	$\psi$ (deg)		& $50\pm3$		& $40\pm3$		& $52\pm9$ 		& $42\pm6$	& $61\pm2$\tablenotemark{***}	& $60\pm2$\tablenotemark{***}\\
	impact parameter	& $0.29\pm0.02$ & $0.26\pm0.02$	& $0.29\pm0.01$ & $0.25\pm0.02$	& $0.28\pm0.02$	& $0.24\pm0.02$\\
	stellar oblateness	& $0.022\pm0.001$ & $0.027\pm0.002$ & $0.02\pm0.02$ & $0.022\pm0.006$ & $0.014\pm0.002$ & $0.018\pm0.003$\\
	\midrule
	$\chi_{\rm min}^2/{\rm dof}$  & $250/241$	& $249/241$		& $247/240$	&	$245/240$ & $248/241$	&	$246/241$
	\enddata
	\tablecomments{The quoted best-fit values and uncertainties are averages and standard deviations 
	of the best-fit values obtained from $13$ quarters analyzed here.
	The value of $\chi^2_{\rm min}$ is also the average of the minimum $\chi^2$ among quarters.
	}
	\tablenotetext{*}{This should be read as $-58\pdeg6\pm2\pdeg0$ for the solution 
	with $\cos i_{\rm orb}<0$ discussed in this table.}
	\tablenotetext{**}{Measured from the transit epoch $t_0({\rm BJD-2454833})=120.566\pm0.001$
	obtained with the transit model without gravity darkening.}
	\tablenotetext{***}{For $\lambda$ in the joint solutions, we quote the uncertainty in the constraint
	from the Doppler tomography.
	This is because the value of $\lambda$ is completely determined by this
	constraint and its standard deviation (several $0.1^\circ$) is not a good measure 
	of the actual uncertainty.
	Accordingly, the quoted uncertainty in $\psi$ is also increased
	by taking a quadratic sum of its standard deviation and the additional scatter coming from 
	the uncertainty of $2^\circ$ in $\lambda$.
	}
	\label{tab:koi13_gdfit}
\end{deluxetable*}
\section{Spin--orbit precession in the Kepler-13A system}\label{sec:precession}
The shape of Kepler-13Ab's transit is known to exhibit a long-term variation, 
which is likely due to the spin--orbit precession induced 
by the quadrupole moment of the rapidly rotating host star 
\citep{2012MNRAS.421L.122S, 2014MNRAS.437.1045S}.
Indeed, we find the monotonic decrease in $|\cos i_{\rm orb}|$ from the quarter-by-quarter analysis
in Section \ref{sec:koi13};
the constant-value model is rejected at the $p$-value of $0.5\%$ for this parameter
using a simple $\chi^2$ test.
On the other hand, the other model parameters are found to be consistent with the constant value
using the same criterion.
Therefore, our analysis confirms that the observed TDVs are actually due to the variation in $\cos i_{\rm orb}$,\footnote{Note that, in \citet{2012MNRAS.421L.122S}, 
the degeneracy between $a/R_\star$ (or $\rho_\star$) and $\cos i_{\rm orb}$ was not solved.}
further supporting the precession scenario with the more realistic model of the asymmetric transit light curve.

In this section, we further examine this scenario with the gravity-darkened transit model. 
Unlike the above previous studies \citep{2012MNRAS.421L.122S, 2014MNRAS.437.1045S}
that focused on $i_{\rm orb}$, 
the gravity-darkened model allows us to additionally study the (non-)variations in
the other two angles, $\lambda$ and $i_\star$,
which should also be induced if the system is precessing.\footnote{
If either of the angular momenta of the stellar spin or the orbital motion dominates,
$i_{\rm orb}$ or $i_\star$ is almost constant.
In the Kepler-13A system, the two angular momenta have comparable magnitudes
and so all three angles modulate due to the precession.
A similar case, the PTFO 8-8695 system, has been studied by \citet{2013ApJ...774...53B} 
and S. Kamiaka et al. (2015, in preparation).}
By fitting the analytic precession model to the time series of 
$\cos i_{\rm orb}$, $\lambda$, and $i_\star$ obtained from the light curves,
we constrain the stellar quadrupole moment $J_2$ and its moment of inertia coefficient $\mathbb{C}$.
On the basis of these constraints, we predict the future evolution of the system configuration
and argue that the follow-up observations of such a long-term modulation
can distinguish the light-curve and joint solutions discussed in Section \ref{sec:koi13}.
In the following, we mainly discuss the results obtained with the stellar parameters from S14,
though the conclusions remain the same for the B11 parameters.

\subsection{Model parameters from each transit}\label{ssec:tpars}
\begin{figure*}
	\centering
	\includegraphics[width=13cm, clip]{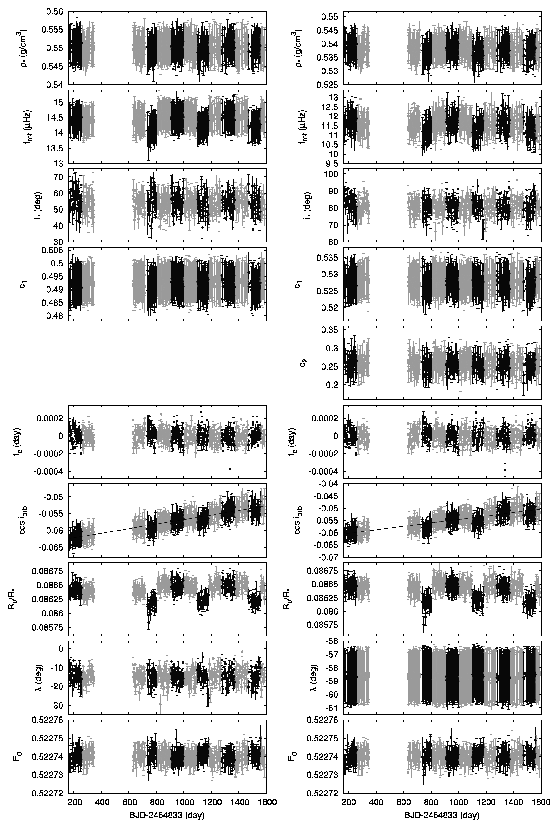}
	\caption{Best-fit model parameters from each transit. The left panels are the results 
	for the light-curve solution with $c_2=0$, while the right ones are for the joint solution.
	Errors are from the outputs of the {\tt cmpfit} package.
	Parameters from even quarters (2, 8, 10, 12, 14, and 16) are shown in black, 
	while those from odd quarters (3, 7, 9, 11, 13, 15, and 17) are in gray.
	For the times of inferior conjunctions, $t_c$, the residuals of the linear fit (i.e., TTVs) are 
	plotted for clarity.
	Solid lines in $\cos i_{\rm orb}$ panels are the best-fit linear models.}
	\label{fig:tpars}
\end{figure*}

To examine the temporal variations in $\cos i_{\rm orb}$, $i_\star$, and $\lambda$,
we fit individual transit light curves, rather than the phase-folded ones, for these parameters.
We use the same two models (``light-curve solution" with $c_2=0$ and 
``joint solution" with $c_2$ fitted) as discussed in Section \ref{sec:koi13}.
In order not to underestimate the errors in the three angles,
we fit all the other model parameters, $\rho_\star$, $c_1$, $c_2$ (for the joint model), $t_c$,
$R_{\rm p}/R_\star$, $f_{\rm rot}$, and $F_0$ as well,
which should not vary temporally in our model.
Using the best values in Table \ref{tab:koi13_gdfit},
we impose the constraints on these parameters except for $t_c$ and $F_0$, 
through the second term of Equation (\ref{eq:chi2}).
In fitting much noisier individual transits,
this prescription assures that the parameters converge to the values consistent 
with those from the phase-folded light curves, while preserving their differences 
from transit to transit.
We also discard the transits for which fit does not converge
due to the data gaps and/or flare-like brightening features sometimes found in the light curves.
The resulting sequences of the transit parameters are plotted in Figure \ref{fig:tpars}. 

As mentioned above, we again find the clear linear trend in $\cos i_{\rm orb}$ from individual transits.
We fit the linear model to the time series of $\cos i_{\rm orb}$ 
using a Markov Chain Monte Carlo (MCMC) algorithm
and obtain the rates of change in the upper part of Table \ref{tab:fit_angles}.
Here we only report the slopes for absolute values of $\cos i_{\rm orb}$
because its actual sign depends on the sign of $\cos i_\star$,
which can never be determined with the current observations 
(we arbitrarily choose $\cos i_\star>0$ in this paper, as discussed after Equation \ref{eq:psi}).
Comparing the light-curve solution and joint solution,
we find that the rate of $|\cos i_{\rm orb}|$ change is insensitive to $\lambda$ or $c_2$
because $|\cos i_{\rm orb}|$ is mainly determined from the transit duration.
With $a/R_\star$ calculated from $\rho_\star$, our value for $\mathrm{d} |\cos i_{\rm orb}|/\mathrm{d}t$ 
is found to be consistent with 
$\mathrm{d}b/\mathrm{d}t = (-4.4\pm1.2)\times10^{-5}\,\mathrm{day}^{-1}$ by \citet{2012MNRAS.421L.122S},
but our constraint is several times better.

Figure \ref{fig:tpars} also shows the abrupt systematic changes in $R_{\rm p}/R_\star$.
These changes occur exactly in phase with the border of different quarters 
indicated with different colors (black and gray).
For this reason, they are unlikely to be of physical origin,
but are probably due to the seasonal transit depth variations 
similar to those reported by \citet{2013ApJ...774L..19V} for HAT-P-7.
In addition, some of the parameters (most notably $\rho_\star$ and $f_{\rm rot}$)
show the long-term modulation of the period $\sim 400\,\mathrm{days}$.
Origins of these systematics are beyond the scope of this paper,
and they are just treated as the additional scatter in the data.
\begin{deluxetable*}{lcccc}
	\tabletypesize{\small}
	\tablewidth{0pt}
	\tablecolumns{5}
	\tablecaption{Results of the precession model fit 
	to $\cos i_{\rm orb}$, $i_\star$, and $\lambda$ from each transit}
	\tablehead{
	& \multicolumn{2}{c}{light-curve solution ($c_2=0$)}
	& \multicolumn{2}{c}{joint solution ($c_2$ fitted)}\vspace{0.1cm}\\
	\colhead{Ref. for $F_{\rm c}$, $v\sin i_\star$, $M_\star$, $T_{\star,\rm pole}$}  
	& \colhead{B11}	& \colhead{S14}	& \colhead{B11}	& \colhead{S14}
	}
	\startdata
	\multicolumn{2}{l}{({\it Linear fit to $\cos i_{\rm orb}$})}\\
	$|\cos i_{\rm orb}|$\tablenotemark{*}
	&	$0.0668\pm0.0001$	&	$0.0581\pm0.0001$ 	&	$0.0658\pm0.0001$	&	$0.0560\pm0.0002$\\
	$\frac{\mathrm{d}|\cos i_{\rm orb}|}{\mathrm{d}t}\,(\mathrm{day}^{-1})$
	&	$(-5.9\pm0.3)\times10^{-6}$	&	$(-6.8\pm0.3)\times10^{-6}$	&	$(-6.0\pm0.3)\times10^{-6}$	&	$(-7.0\pm0.4)\times10^{-6}$\\
	\vspace{-0.2cm}\\
	\multicolumn{2}{l}{({\it Precession model fit to $\cos i_{\rm orb}$, $i_\star$, and $\lambda$})}\\
	$\rho_\star$, $f_{\rm rot}$, $P$		
		&\multicolumn{4}{c}{Same as Table \ref{tab:koi13_gdfit} (priors $=$ posteriors)}\\
	$\cos i_{\rm orb}$\tablenotemark{*}		
		& $-0.0668\pm0.0001$	& $-0.0581\pm0.0001$	& $-0.0658\pm0.0001$	&	$-0.0560\pm0.0002$\\
	$i_\star$ (deg)\tablenotemark{*}	
		& $44.7\pm0.3$		& $54.2\pm0.3$	& $72.8\pm0.3$	& $81.8\pm0.2$\\
	$\lambda$ (deg)\tablenotemark{*}		
		& $-20.1\pm0.2$ 	& $-13.9\pm0.1$	& $-58.65\pm0.09$	& $-58.62\pm0.09$\\
	$M_{\rm p}/M_\star$\tablenotemark{**}	
		& $(3.4\pm0.8)\times10^{-3}$	& $(2.8\pm0.8)\times10^{-3}$	& $(4.1\pm0.8)\times10^{-3}$	&	$(4.0\pm0.8)\times10^{-3}$\\
	$\mathbb{C}$\tablenotemark{***}		
		& $0.09\pm0.02$ & $0.10\pm0.02$	& $0.08\pm0.02$	& $0.08\pm0.02$\\
	$J_2$								
		& $(1.44\pm0.07)\times10^{-4}$ & $(1.66\pm0.08)\times10^{-4}$ & $(5.6\pm0.3)\times10^{-5}$ & $(6.1\pm0.3)\times10^{-5}$\\
	\vspace{-0.2cm}\\
	\multicolumn{2}{l}{({\it Derived from the precession model})}\\
	Precession period (yr) & $(5.7\pm0.4)\times10^2$ &  $(4.3\pm0.3)\times10^2$ & $(1.6\pm0.2)\times10^3$ 	&	$(1.5\pm0.2)\times10^3$\\
	$L/S$				 & $0.36_{-0.09}^{+0.11}$	& $0.25\pm0.07$	& $0.65_{-0.17}^{+0.24}$	&	$0.54_{-0.14}^{+0.19}$	
	\enddata
	\tablecomments{The quoted values and uncertainties are $50$, $15.87$, and $84.13$ 
	percentiles of the marginalized MCMC posteriors.}
	\tablenotetext{*}{Value at $\mathrm{BJD}=2455633=2454833+800$.}
	\tablenotetext{**}{Gaussian prior $M_{\rm p}/M_\star=(4.2\pm0.8)\times10^{-3}$ is imposed. 
	The value is based on the average and standard deviation of the results by
	S14, \citet{2014arXiv1407.2245E}, and \citet{2014arXiv1407.2361F}.}
	\tablenotetext{***}{Gaussian prior $\mathbb{C}=0.0776\pm0.0200$ is imposed. The central value is from
	the result for $n=3$ polytrope by \citet{2012MNRAS.421L.122S}
	and the width is chosen to enclose that of the Sun.}
	\label{tab:fit_angles}
\end{deluxetable*}
\subsection{Fit to the observed angles and future prediction}
Among the observed time series of transit parameters in Figure \ref{fig:tpars},
those of $\cos i_{\rm orb}$, $\lambda$, and $i_\star$ are fitted using an MCMC algorithm
to observationally constrain $J_2$ and $\mathbb{C}$.
We utilize the same analytic precession model as in \citet{2013ApJ...774...53B},
which constitutes an analytic solution of the secular equations of motion 
derived by \citet{2009Icar..201..750B}.
In this model, the orbital and spin angular momenta precess around the total angular momentum 
at the same angular rate given by
\begin{equation}
	\dot \Omega = \dot \Omega_{\rm p} \sqrt{\left(\frac{L}{S}+\cos \psi \right)^2 + \sin^2 \psi},
	\label{eq:domega}
\end{equation}
where $\dot \Omega_{\rm p}$ is the precession rate
of the orbital angular momentum around the stellar spin, and explicitly given by
\begin{equation}
	\dot \Omega_{\rm p} = - \frac{3}{2} J_2 \frac{2\pi}{P} \cos \psi \left(\frac{R_\star}{a}\right)^2
	\label{eq:domega_p}
\end{equation}
with $J_2$ being the stellar quadrupole moment. In the Kepler-13A system, 
the spin angular momentum, $S$, is comparable to the orbital one, $L$, 
owing to the small semi-major axis and rapid stellar rotation.
As a consequence, $\dot \Omega$ also depends on the ratio of the two,
\begin{equation}
	\frac{L}{S} = \frac{1}{\mathbb{C}} \frac{M_{\rm p}}{M_\star} \frac{1}{P f_{\rm rot}} \left(\frac{a}{R_\star}\right)^2,
	\label{eq:ls}
\end{equation}
where $\mathbb{C}$ is the moment of inertia coefficient of the host star.
Thus, the independent model parameters are 
$\rho_\star$, $f_{\rm rot}$, $J_2$, $\mathbb{C}$, $P$, $M_{\rm p}/M_\star$,
and three angles $\cos i_{\rm orb}$, $\lambda$, $i_\star$ at some epoch (here taken to be 
$\mathrm{BJD}=2454833+800$).
We do not relate $J_2$ to the other parameters like the stellar oblateness as done in \citet{2013ApJ...774...53B}.

To realistically evaluate the credible intervals of $J_2$ and $\mathbb{C}$ by marginalization,
uncertainties in $\rho_\star$, $f_{\rm rot}$, $P$, and $M_{\rm p}/M_\star$ should also be taken into account.
However, these parameters are not well determined from the data of $\cos i_{\rm orb}$, $\lambda$, and $i_\star$.
Thus, they are floated with the following Gaussian priors.
The first three are assigned the same central values and widths
as in Table \ref{tab:koi13_gdfit}.
For the mass ratio, we take the mean and standard deviation of the results reported by
S14, \citet{2014arXiv1407.2245E}, and \citet{2014arXiv1407.2361F},
which come from the amplitudes of the ellipsoidal variation and Doppler beaming.
We also impose the Gaussian prior on $\mathbb{C}$
centered on $0.0776$ \citep[the value for $n=3$ polytrope by][]{2012MNRAS.421L.122S}
and with the width of $0.02$,
which is chosen to enclose the solar value, $0.059$.

\begin{figure*}
	\centering
	\includegraphics[width=14cm]{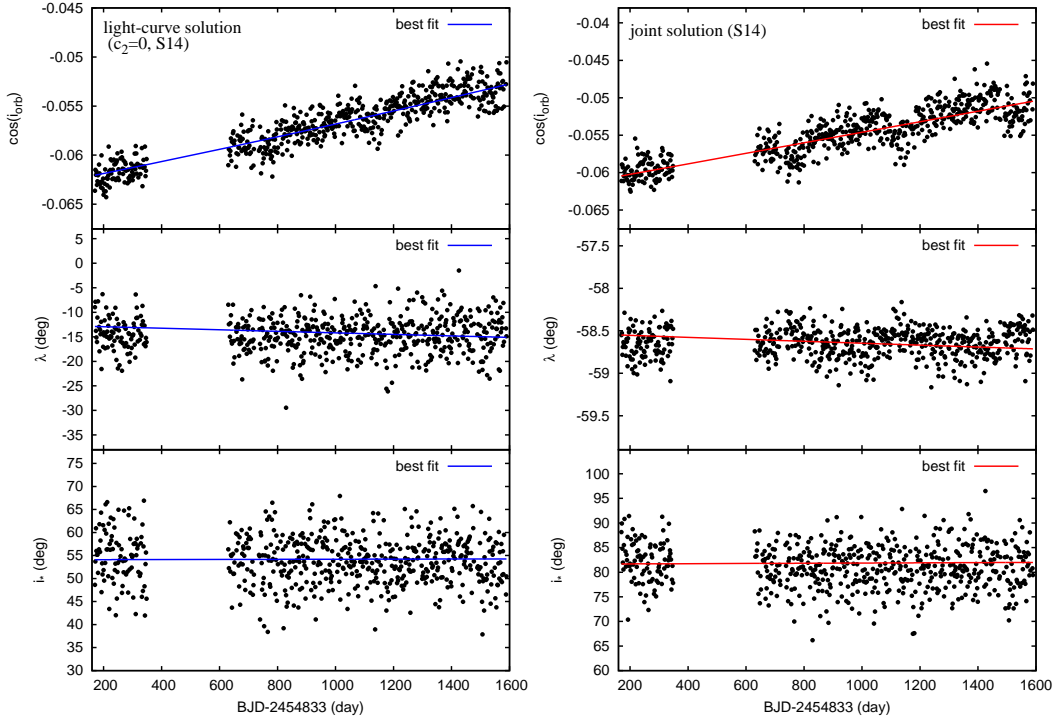}
	\caption{Simultaneous fit to the observed $\cos i_{\rm orb}$, $\lambda$, and $i_\star$. 
	(Left) light-curve solution with $c_2=0$. (Right) joint solution.
	Black points are from the light-curve fit (same as Figure \ref{fig:tpars}),
	and colored solid lines denote the best-fit precession models,
	which are {\it not} the linear fits.}
	\label{fig:fit_angles}
\end{figure*}

The constraints from the MCMC fit are summarized in 
the middle and bottom parts of Table \ref{tab:fit_angles}
and the best-fit models are plotted with the solid lines in Figure \ref{fig:fit_angles}.
Basically, the precession model is compatible with the observations 
both for the light-curve solution and the joint solution.
The value of $J_2$ and the corresponding precession period, however,
are different by a factor of a few, in spite of the similar observed slopes in $\cos i_{\rm orb}$.
While $J_2=(1.66\pm0.08)\times10^{-4}$ for the light-curve solution is consistent with the earlier estimate by \citet{2012MNRAS.421L.122S},
$J_2=(2.1\pm0.6)\times10^{-4}$ from observed TDVs and $J_2=1.7\times10^{-4}$ from the stellar model, 
the joint solution yields a smaller value, $J_2=(6.1\pm0.3)\times10^{-5}$.

\begin{figure*}
	\centering
	\includegraphics[width=14cm,clip]{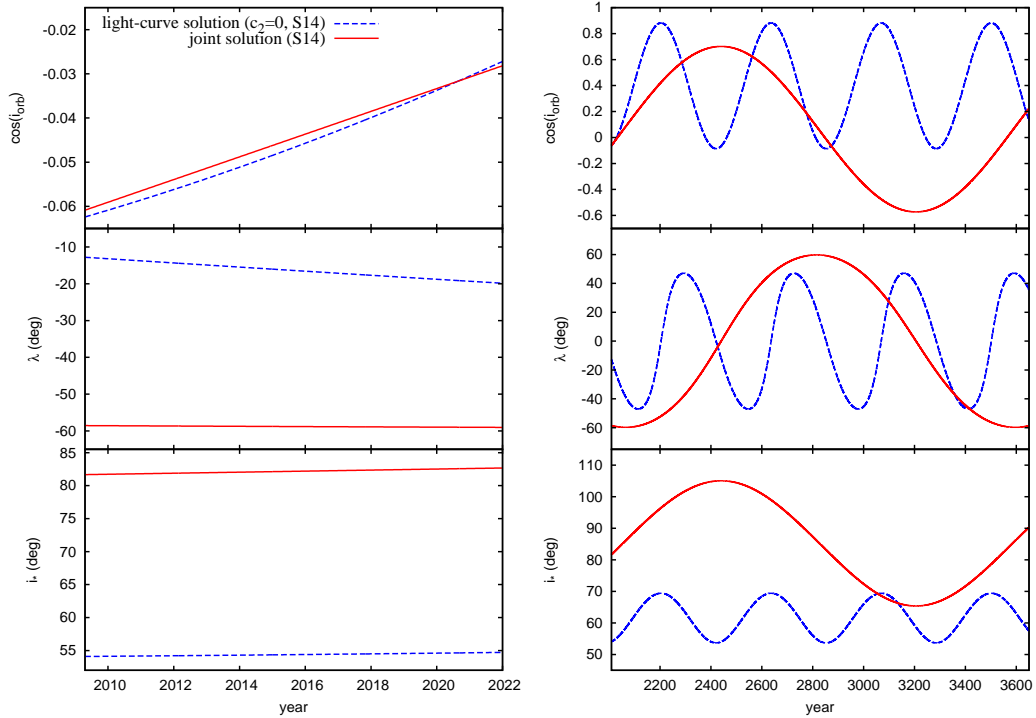}
	\caption{Future evolutions of $\cos i_{\rm orb}$, $\lambda$, and $i_\star$ predicted for the best-fit
	models in Table \ref{tab:fit_angles} and Figure \ref{fig:fit_angles}.
	From top to bottom, the evolutions of $\cos i_{\rm orb}$, $\lambda$, and $i_\star$ are plotted for 
	the solution from the light-curve alone (blue dashed line) and joint solution (red solid line) 
	obtained with the S14 stellar parameters.
	The left panels show the short-term ($\sim 14\,\mathrm{yr}$, until 2022) behavior, 
	while the right ones are for the long-term ($\sim1600\,\mathrm{yr}$) variation.}
	\label{fig:angles_evolution}
\end{figure*}

The difference comes from the different three-dimensional architectures of the system
described by the two solutions.
Since all of $\cos i_{\rm orb}$, $\lambda$, and $i_\star$ are constrained from the gravity-darkened light curves,
relative configuration of the stellar spin and orbital angular momenta are completely specified
in three dimensions.
This means that the {\it phase} of the precession during the {\it Kepler} mission,
which corresponds to the left end in the right column of Figure \ref{fig:angles_evolution},
is observationally constrained;
from the top panel, we find that $\cos i_{\rm orb}$ 
is closer to the bottom of the sine curve for the light-curve solution (blue dashed line),
while that for the joint solution (red solid line) resides in the phase of a rapid increase.
For this reason, a larger precession rate (i.e., shorter precession period) 
is required for the light-curve solution to match the observed change in $\cos i_{\rm orb}$.
According to Equations (\ref{eq:domega}) and (\ref{eq:domega_p}), the larger precession rate can be achieved by increasing 
either $J_2$ or $L/S$. 
However, the larger precession rate also induces faster variations in $\lambda$ and $i_\star$,
contradicting their almost constant observed values (middle and bottom panels in Figure \ref{fig:fit_angles}).
The only way to mitigate this conflict is to make $J_2$ larger (i.e., increase the precession rate)
while keeping $L/S$ small, making it more difficult to move stellar spin axis.
With Equation (\ref{eq:ls}), this explains why $M_{\rm p}/M_\star$ is smaller and $\mathbb{C}$ is larger for the light-curve solution
than for the joint solution in Table \ref{tab:fit_angles}. 
Accordingly, the bottom panel of the right column in Figure \ref{fig:angles_evolution} exhibits
the smaller precession amplitude for $i_\star$ in the former solution (blue dashed line) 
than the latter (red solid line).


The approximately three times difference in the precession period would be apparent even 
on the short time scale (left column in Figure \ref{fig:angles_evolution}).
As shown in the middle panel, as large as $\sim 10^\circ$ change in $\lambda$ is expected 
within the next $\sim 10\,\mathrm{yr}$ for the light-curve solution,
which may well be detectable given the current precision of the spin--orbit angle measurement
(nominally down to a few degrees).
On the other hand, $\lambda$ for the joint solution is almost constant.
From this point of view, the joint solution may slightly be favored even with the current data,
because the nearly-constant values observed for $\lambda$ and $i_\star$ are more natural for
the joint solution than for the light-curve one, for the reasons discussed in the previous paragraph.
This indication also manifests itself in the fact that the resulting $M_{\rm p}/M_\star$ and $\mathbb{C}$ 
better agree with our prior knowledge in the joint solution.

The more decisive conclusion will be obtained with the future follow-up observations of $\lambda$
using Doppler tomography, as well as the transit duration observations to better constrain $\cos i_{\rm orb}$,
and hence the precession rate.
If our joint solution is correct, variations in $\lambda$ will not be detected in near future.
On the other hand, if the light-curve solution is actually correct and $\lambda$ from the Doppler tomography
is somehow systematically biased, $\lambda$ should change;
this {\it temporal variation} would be observable with the Doppler tomography even if it were biased.
Or, it may even turn out that the precession scenario is wrong.
In any case, tracking the future evolution of the system configuration 
can be used for an independent test of our solution,
not to mention for better constraining stellar internal structure via $J_2$ and $\mathbb{C}$,
for which few observational constraints have been obtained.

\section{Anomaly in the Transit Light Curve of HAT-P-7}\label{sec:hatp7}
Armed with the methodology established using the distinct anomaly in Kepler-13A
(Section \ref{sec:koi13}),
we discuss another, more subtle anomaly in this section.
Here the methodology is further extended 
to include the information from asteroseismology as well as from the RM effect,
and applied to an F-type star.

It has been pointed out in several studies that the transit light curve of HAT-P-7
exhibits a small anomaly of $\mathcal{O}(10^{-5})$.
\citet{2013ApJ...764L..22M}, who reported this anomaly first, attributed it to the 
local spot-like gravity darkening induced by the gravity of 
the Jupiter-mass companion HAT-P-7b.
They ruled out the gravity darkening of stellar rotational origin
on the basis of the inspection that the anomaly is localized in a part of the transit.
Later analyses with more data \citep[e.g.,][]{2013ApJ...772...51E, 2013ApJ...774L..19V, 2014arXiv1407.2245E, 2014PASJ...66...94B}, 
however, have shown that the anomaly is seemingly correlated over the whole transit duration,
as in the top panel of Figure \ref{fig:hatp7_best}.
Moreover, the amplitude of the observed anomaly may be too large to be explained by the spot scenario.
According to \citet{2012ApJ...751..112J}, the planet's gravity induces the surface temperature variation of ``a few $0.1\,\mathrm{K}$,"
which leads to the surface brightness variation of $\Delta F \sim \mathrm{several}\ 100\,\mathrm{ppm}$.
If a planet crosses over a spot fainter by $\Delta F$ than the other part of the stellar disk, amplitude of the expected 
anomaly in the {\it relative flux} is about $\Delta F \times (R_{\rm p}/R_\star)^2 \sim \mathcal{O}(\mathrm{ppm})$, which is order-of-magnitude
smaller than the observed one.
We therefore analyze this anomaly assuming that it is originated from the gravity darkening induced by stellar rotation,
whose effect should not be localized but manifest during the whole transit duration.

Unlike the case of Kepler-13A, anomaly in the transit light curve is not clear 
on a quarter-by-quarter basis for HAT-P-7.
In addition, no TTVs/TDVs have been detected for this planet.
For these reasons, we deal with the light curve obtained by folding all the available 
SC, PDCSAP fluxes (Q0--17) processed as described in Section \ref{ssec:method_data}.
We use the spectroscopic constraint $v\sin i_\star=3.8\pm1.5\,\mathrm{km\,s^{-1}}$ throughout this section.
This value is based on \citet{2008ApJ...680.1450P}, though its error bar is enlarged to take into account 
other estimates for this quantity that give slightly different values \citep[e.g.,][]{2009ApJ...703L..99W}.

\subsection{Robustness of the Observed Anomaly}\label{ssec:hatp7_robustness}
\begin{figure*}[htbp]
	\centering
	\includegraphics[width=16cm, clip]{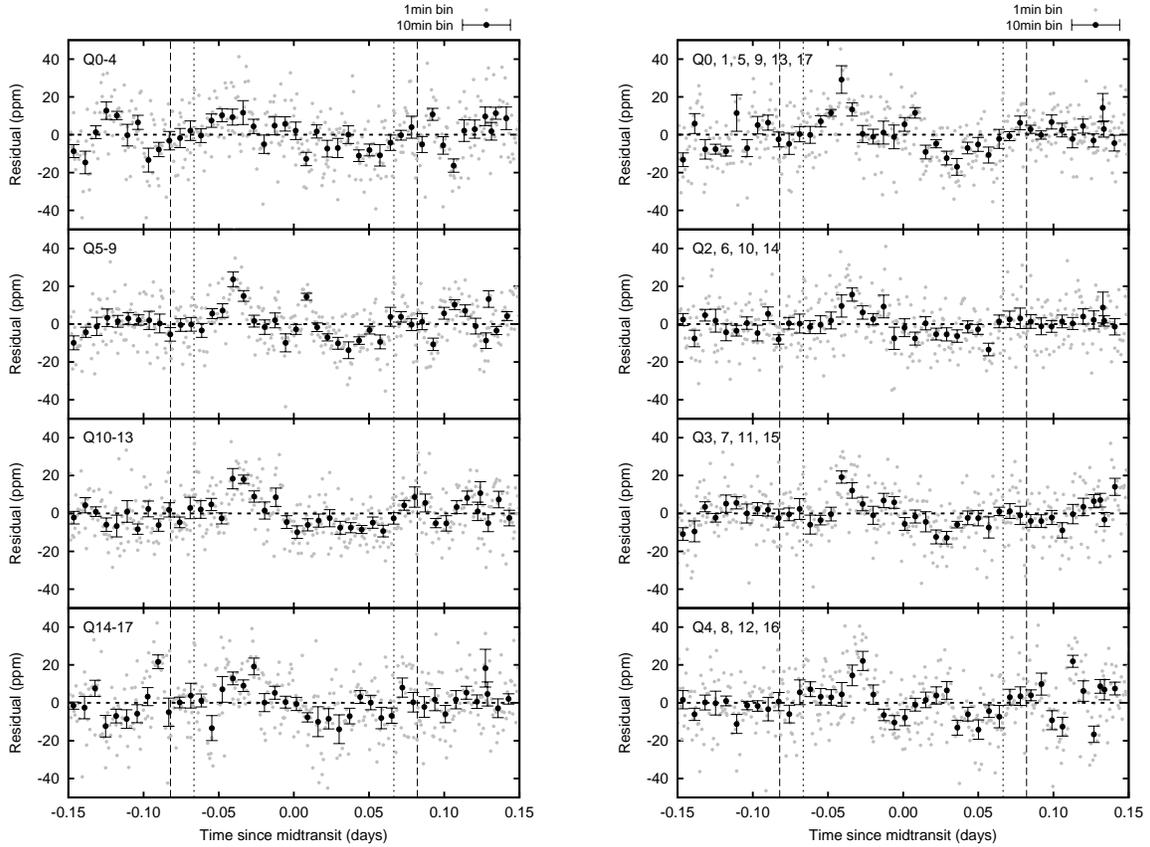}
	\caption{Robustness of the detected anomaly. 
	Residuals of the model fits (without gravity darkening) to the phase-folded transit light curves are plotted.
	Gray dots are residuals for the one-minute binned data, 
	and black ones with error bars are the residuals averaged into ten-minutes bins.
	Vertical dashed and dotted lines correspond to the beginnings and ends of the ingress and egress.
	(Left column) Transits folded over different epochs. From top to bottom, 
	light curves from Quarters 0--4, 5--9, 10--13, and 14--17 are folded.
	(Right column) Transits grouped by the CCD module used to observe the target.
	From top to bottom, light curves taken with the modules 17, 19, 9, and 7 are folded.}
	\label{fig:seasons}
\end{figure*}

If the observed anomaly is really due to gravity darkening, it should be persistent over all observation span.
It is important to confirm the property 
because \citet{2013ApJ...764L..22M} only reported the bump before the mid-transit time.
Thus, we divide the transits into four consecutive groups (Q0--4, 5--9, 10--13, 14--17),
phase-fold and fit each of them with the model without gravity darkening separately,
and examine the shapes of the residuals.
Although fewer numbers of transits lead to noisier phase-folded light curves,
ten-minute binned residuals in the left column of Figure \ref{fig:seasons} 
exhibit a similar feature 
(brightening before mid-transit and dimming after it) in every span of data.

Besides, \citet{2013ApJ...774L..19V} reported seasonal variation in the transit depth 
depending on the quarter,
which is reproduced in our analysis with Q0--17 data.\footnote{
We also reported a similar phenomenon in Kepler-13A; see Section \ref{ssec:tpars}
and Figure \ref{fig:tpars}.}
To confirm that the anomaly is not an artifact related to this seasonal variation,
we also perform a similar analysis as above but this time grouping the transits 
that have similar depths.
As shown in the right column of Figure \ref{fig:seasons}, we find that the 
same feature is apparent regardless of the season 
and the anomaly is not affected by the systematic depth variation.
For this reason, along with its unconstrained origin, we do not try to make corrections for this systematic
in the following analyses.
  
\subsection{Results}\label{ssec:hatp7_results}
\begin{figure*}
	\centering
	\includegraphics[width=12cm, bb=50 50 280 302]{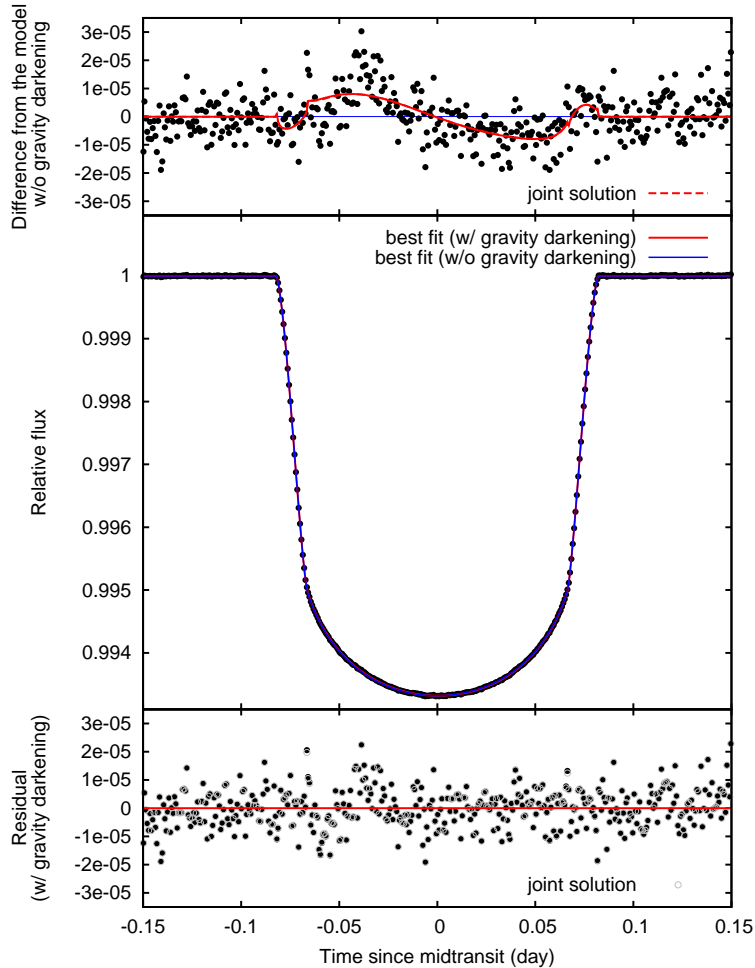}
	\caption{Fitting the gravity-darkened model to the phase-folded transit of HAT-P-7b.
	The meanings of the symbols are the same as those in Figure \ref{fig:koi13_best_q2},
	but this time the joint solution incorporates the constraints on $\lambda$ from the RM measurement 
	and on $f_{\rm rot}$ from asteroseismology.
	The light-curve solution and joint solution are almost indistinguishable in this case,
	as expected from the similar values of $\chi^2$.}
	\label{fig:hatp7_best}
\end{figure*}

As in Section \ref{sec:koi13}, we consider both light-curve solution and joint solution
that takes into account the constraints from other observations.
First, the light-curve solution is obtained with $c_2$ fixed to be zero
(Figure \ref{fig:hatp7_best}, second and third columns in Table \ref{tab:hatp7}).
In this case, we find two solutions with different signs of $\cos i_{\rm orb}$,
which are indistinguishable in terms of the minimum $\chi^2$.\footnote{
The existence of the two solutions in this case should be distinguished 
from the degeneracy intrinsic to the gravity-darkening method.
For each of the two solution listed here, there additionally exists the model that yields exactly the
same light curve, where $\cos i_{\rm orb}$ is replaced with $-\cos i_{\rm orb}$ and $\lambda$ with $\pi-\lambda$.
These intrinsically-degenerate solutions are not discussed here because 
they are in any case rejected in the joint solution, where $\lambda$
is constrained by the prior.
This is the same logic as used in the last paragraph of Section \ref{ssec:koi13_b11}.
}
The values quoted in Table \ref{tab:hatp7} are the median, 15.87, and 84.13 percentiles of the MCMC posteriors
sampled with {\tt emcee} \citep{2013PASP..125..306F}.\footnote{
We also applied the residual permutation method described in \citet{2009ApJ...693..794W}
for another estimate of the parameter uncertainties,
and confirmed that they are not significantly affected by the correlated noise component.}
Our model reasonably reproduces the global feature of the anomaly (positive before the mid transit and negative after it), yielding $\Delta \chi^2 \simeq 166$ for $\sim 420$ degrees of freedom.
We compute the Bayesian information criterion (BIC) for the best-fit models with and without gravity darkening,
and find $\Delta \mathrm{BIC} =129$, which formally indicates
that the gravity-darkened model is strongly favored.

Our solution points to a nearly pole-on configuration with $i_\star\simeq0^\circ$.
This conclusion is consistent with the recent asteroseismic analyses by \citet{2014PASJ...66...94B} and \citet{2014A&A...570A..54L},
but the nominal constraint on $i_\star$ from the gravity-darkened model is much tighter.
On the other hand, $\lambda$ is not constrained very well with the light curve asymmetry alone.
The difficulty is inevitable in the pole-on configuration, where 
the brightness distribution on the stellar disk is almost axisymmetric even in the presence of gravity darkening.
In such a case, $\psi$ is always close to $90^\circ$ regardless of $\lambda$.

One remaining issue regarding our solution is that
the resulting rotation frequency may be too large.
Given the age ($\simeq 2\,\mathrm{Gyr}$) and $B-V$ \citep[$=0.495\pm0.022$;][]{2014A&A...570A..54L}
of the host star, 
the rotation frequency from the light-curve solution, 
$f_{\rm rot}=7.7\pm0.2\,\mathrm{\mu Hz}$ 
(equivalent to $P_{\rm rot} \simeq 1.5\,\mathrm{days}$),
is consistent with the gyrochronology relation by \citet{2009ApJ...695..679M};
see Section 6 of \citet{2014A&A...570A..54L}.
However, our value of $f_{\rm rot}$ is much larger than those from asteroseismology, 
$0.70_{-0.43}^{+1.02}\,\mathrm{\mu Hz}$ \citep[$68\%$ credible interval by][]{2014PASJ...66...94B}
and $<0.8748\,\mathrm{\mu Hz}$ \citep[$1\sigma$ upper limit by][]{2014A&A...570A..54L}.
In fact, the prior used in these analyses, $|f_{\rm rot}|<8\,\mathrm{\mu Hz}$,
does not fully cover the range we investigate here with the gravity-darkened light curve.
Still, the discrepancy is only weakly reduced even with
the new analysis adopting the prior range extended up to $17\,\mathrm{\mu Hz}$, 
which yields $f_{\rm rot}=0.82_{-0.50}^{+2.02}\,\mathrm{\mu Hz}$
as the $68\%$ credible interval 
\citep[by courtesy of Othman Benomar; see also][]{2014PASJ...66...94B}.

To examine if the gravity-darkened model is compatible with the seismic analysis, 
we then search for a joint solution including the constraints both from the RM measurement and asteroseismology.
From the RM effect, we incorporate the constraint $\lambda=172^\circ \pm 32^\circ$,
which comes from the average and standard 
deviation of the analyses for the three different radial velocity data \citep{2014PASJ...66...94B}.
From asteroseismology, we adopt the above updated posterior for $f_{\rm rot}$ as the prior,
and performed an MCMC sampling with {\tt emcee}.
To properly take into account the uncertainty from the limb-darkening profile, 
$c_2$ is also floated.
The resulting credible intervals are summarized in the fourth and fifth columns in Table \ref{tab:hatp7},
and the model that maximizes the likelihood multiplied by the prior on $f_{\rm rot}$ is plotted with a dashed 
line in Figure \ref{fig:hatp7_best}.
We again find two equally good solutions, 
both of which indicate nearly pole-on configurations
with slightly prograde and retrograde orbits, $\psi=101^\circ\pm2^\circ$ and $\psi=87^\circ\pm2^\circ$.
Although the resulting $f_{\rm rot}$ still prefers a higher rotation rate
than that from asteroseismology,
their difference is now mitigated to the $2\sigma$ level;
we construct the probability distribution for $\Delta f_{\rm rot}$,
$f_{\rm rot}$ from out joint analysis minus $f_{\rm rot}$ from asteroseismology, using their posteriors
and find its $2\sigma$ credible region as $\Delta f_{\rm rot} = 4.9_{-5.0}^{+4.0}\,\mu\mathrm{Hz}$.
We argue that the level of discrepancy is acceptable, 
considering that the rotational mode splitting is not clearly detected 
in the power spectrum of HAT-P-7's light curves.

Finally, it is also worth considering the case where $\beta \neq 0.25$,
given the unconstrained nature of the gravity darkening in F dwarfs.
Smaller values of $\beta\sim0.08$ are usually expected for solar-like stars with convective envelopes
\citep[e.g.,][]{1967ZA.....65...89L, 1998A&AS..131..395C}, 
while \citet{2011A&A...533A..43E} and \citet{2012A&A...547A..32E} 
argue that $\beta$ is close to $0.25$ in the limit of slow rotation under several assumptions.
We repeat the above joint analysis floating $\beta$ with the prior uniform between $0$ and $0.3$,
and obtain $\beta=0.26_{-0.05}^{+0.03}$ for both solutions in Table \ref{tab:hatp7}.
On the one hand,
the fact may support the claims by \citet{2011A&A...533A..43E} and \citet{2012A&A...547A..32E};
on the other hand, it may simply indicate some incompleteness in our gravity-darkening model,
as also suggested by the tension in $f_{\rm rot}$ and the still correlated residuals 
before the mid transit (bottom panel of Figure \ref{fig:hatp7_best}). 
Indeed, if $\beta=0.08$ is adopted, we find that even higher rotation rate ($>10\,\mathrm{\mu Hz}$) 
is favored, making the discrepancy with asteroseismology more serious.
Although the validity of $\beta$ we obtain is beyond the scope of this paper,
we note that our conclusion for a pole-on orbit is robust against the adopted value of $\beta$;
in both analyses where $\beta$ is fitted and $\beta$ is fixed to be $0.08$,
the constraints on $\psi$ differ less than $1\sigma$
from the results in Table \ref{tab:hatp7}.
\begin{deluxetable*}{lcccc}
	\tabletypesize{\small}
	\tablewidth{0pt}
	\tablecolumns{5}
	\tablecaption{Results for the transit of HAT-P-7${\rm \b}$}
	\tablehead{
	& \multicolumn{2}{c}{light-curve solution ($c_2=0$)}
	& \multicolumn{2}{c}{joint solution ($c_2$ fitted)}\vspace{0.1cm}\\
	& \colhead{Solution 1} & \colhead{Solution 2} & \colhead{Solution 1} & \colhead{Solution 2}
	}
	\startdata
	\multicolumn{1}{l}{({\it Constraints})}\\
	$v\sin i_\star$ ($\mathrm{km\,s^{-1}}$) & $3.8\pm1.5$ & $3.8\pm1.5$ & $3.8\pm1.5$ & $3.8\pm1.5$\\
	$\lambda$		&	$\cdots$		& $\cdots$ &	$172\pm32$	&	$172\pm32$\\
	\vspace{-0.2cm}\\
	({\it Fitted Parameters})\\
	$M_\star$ ($M_\odot$)	& $1.59$ (fixed) & $1.59$ (fixed) & $1.59$ (fixed) & $1.59$ (fixed)\\
	$T_{\star, \rm pole}$ (K)	& $6310$ (fixed) & $6310$ (fixed) & $6310$ (fixed) & $6310$ (fixed)\\
	$\rho_\star$ ($\gcc$)		& $0.2789\pm0.0006$	 		& $0.2789\pm0.0006$& $0.2790\pm0.0005$ 	& $0.2784\pm0.0005$ \\
	$c_1$					& $0.498\pm0.003$			& $0.498\pm0.003$	& $0.507_{-0.016}^{+0.008}$	& $0.508_{-0.015}^{+0.007}$	\\
	$c_2$					& $0$ (fixed)					& $0$ (fixed) 			& $0.07_{-0.12}^{+0.06}$	& $0.08_{-0.11}^{+0.05}$\\
	$t_c$ ($10^{-5}\,\mathrm{day}$)\tablenotemark{*} & $-1.5\pm0.4$ 	& $-1.5\pm0.4$ 		& $-1.6\pm0.4$			& $-1.1\pm0.4$\\
	$P\strut$ (day)					& \multicolumn{4}{c}{\lfill\ $2.204735471$ (fixed)\ \lfill}\\
	$\cos i_{\rm orb}$		& $-0.1195\pm0.0004$		& $0.1195\pm0.0004$ & $-0.1194\pm0.0003$	& $0.1198\pm0.0003$\\
	$R_{\rm p}/R_\star$	& $0.07757_{-0.00009}^{+0.00005}$& $0.07757_{-0.00009}^{+0.00005}$ & $0.07759\pm0.00003$ & $0.07749_{-0.00004}^{+0.00003}$\\
	$F_0\strut$					& \multicolumn{4}{c}{\lfill\ $0.9999998\pm0.0000005$\ \lfill}\\
	$f_{\rm rot}$ ($\mathrm{\mu Hz}$) & $7.7\pm0.2$ 		& $7.7\pm0.2$ 		& $6.1_{-1.7}^{+2.6}$\tablenotemark{**}		& $5.6_{-1.7}^{+2.4}$\tablenotemark{**}\\	
	$i_\star$ (deg)\tablenotemark{***}			& $3.3_{-1.0}^{+1.2}$			& $3.3_{-1.0}^{+1.3}$ 	& $5.3_{-2.0}^{+3.3}$		& $5.3_{-2.1}^{+3.7}$\\
	$\lambda$ (deg) 			& $133_{-88}^{+19}$			& $49_{-21}^{+92}$	& $142_{-16}^{+12}$		& $136_{-22}^{+16}$\\
	$\beta$ 					& $0.25$ (fixed) & $0.25$ (fixed) & $0.25$ (fixed) & $0.25$ (fixed)\\
	\vspace{-0.2cm}\\ 
	({\it Derived Parameters})\\
	$P_{\rm rot}$ (day) & $1.51\pm0.03$ 	& $1.51\pm0.03$		& $1.9_{-0.6}^{+0.7}$	& $2.1_{-0.6}^{+0.9}$\\
	$\psi$ (deg)				& $99_{-4}^{+2}$				& $81_{-2}^{+4}$		& $101\pm2$	& $87\pm2$\\
	impact parameter			& $0.496\pm0.001$			& $0.496\pm0.001$	& $0.496\pm0.001$		& $0.497\pm0.001$\\
	oblateness				& $0.0149\pm0.0006$			& $0.0149\pm0.0007$ & $0.009_{-0.005}^{+0.010}$	& $0.008_{-0.004}^{+0.008}$\\
	\midrule
	$\chi^2_{\rm min}/{\rm dof}$		&	$453/424$		&	$455/424$		& $450/424$		& $451/424$
	\enddata
	\tablecomments{The quoted values and uncertainties are $50$, $15.87$, and $84.13$ 
	percentiles of the marginalized MCMC posteriors.
	For the light-curve solution, $\chi^2_{\rm min}$ is the value of $\chi^2$ computed from Equation (\ref{eq:chi2})
	for the maximum likelihood model.
	Equation (\ref{eq:chi2}) is also used for the joint solution,
	but $\chi^2_{\rm min}$ in this case is computed for the model 
	that maximizes the likelihood function multiplied by the prior on $f_{\rm rot}$.
	}
	\tablenotetext{*}{Measured from the transit epoch $t_0({\rm BJD-2454833})=120.358522\pm0.000005$
	obtained with the transit model without gravity darkening.}
	\tablenotetext{**}{Posterior from the seismic analysis is used as the prior.}
	\tablenotetext{***}{We impose the prior uniform in $\cos i_\star$, rather than in $i_\star$,
	which corresponds to the isotropic distribution for the spin direction.}
	\label{tab:hatp7}
\end{deluxetable*}
\section{Summary}\label{sec:summary}
\subsection{Kepler-13A}
First, we analyze the gravity-darkened transit light curve of Kepler-13A adopting 
the same model and stellar parameters as in the previous study by B11.
We reproduce the spin--orbit angles obtained by B11 with more data
(called ``light-curve solution" in this paper)
and also find that the choice of the stellar mass, stellar effective temperature, $v\sin i_\star$, or contaminated flux
affects $\lambda$ or $i_\star$ by less than about $10^\circ$.
If we fit $c_2=u_1-u_2$ as well as $c_1=u_1+u_2$ in the quadratic limb-darkening law, 
on the other hand, a broader range of the spin--orbit angle is allowed.
In fact, this additional degree of freedom may
explain the discrepancy between the solution by B11 and the Doppler 
tomography result by \citet{2014ApJ...790...30J}.
Our new ``joint solution" includes $i_\star=81^\circ\pm5^\circ$,
$\lambda=-59^\circ\pm2^\circ$,
$\psi=60^\circ\pm2^\circ$, and $P_{\rm rot}=24\pm2\,\mathrm{hr}$.
Although the joint solution is compatible with all of the observations made so far,
introducing additional free parameter $c_2$ is not statistically justified, 
nor is it clear if the best-fit value for $c_2$ is physically plausible.

To examine the above issues from a dynamical point of view,
we also analyze the spin--orbit precession in this system.
By analyzing the light curves from each quarter separately,
we confirm that the variation in $|\cos i_{\rm orb}|$ causes the 
transit duration variations first reported by \citet{2012MNRAS.421L.122S},
with more elaborate model taking into account the gravity darkening.
This variation is consistent with the precession of the stellar spin and orbital angular momenta
around the total angular momentum of the system, induced by the oblateness of the rapidly rotating host star.
We thus fit each transit with the gravity-darkened model to determine 
$\cos i_{\rm orb}$, $\lambda$, and $i_\star$ as a function of time,
and then fit them with the precession model 
to constrain the stellar quadrupole moment $J_2$.
For the light-curve solution and the joint solution,
we respectively find $J_2=(1.66\pm0.08)\times10^{-4}$ and $J_2=(6.1\pm0.3)\times10^{-5}$,
which are different by a factor of a few.
Our results predict detectable variations in $\lambda$ on $10$-yr timescale
for the light-curve solution, while it should be almost constant 
for the joint solution.
The difference suggests that the future follow-up observations 
can be used to confirm or refute the joint solution we proposed,
as well as to improve the constraint on $J_2$.

\subsection{HAT-P-7}
Although the anomaly in the transit light curve is much more subtle compared to Kepler-13Ab,
we confirm that the asymmetric residual (not only the bump reported by \citet{2013ApJ...764L..22M} but also the dip) 
exists continuously in the transits of HAT-P-7b.
Thus, we perform the analysis assuming that the gravity-darkening is a viable explanation for the anomaly.
Gravity-darkened transit model favors a nearly pole-on orbit 
($\psi=101^\circ\pm2^\circ$ or $\psi=87^\circ\pm2^\circ$)
and the gravity-darkening exponent $\beta$ close to $0.25$.
The constraint on $\psi$ is insensitive to the choice of the limb-darkening parameters or the gravity-darkening exponent.

On the other hand, the stellar rotation rate from the gravity-darkening analysis 
is about $2\sigma$ higher than the value from asteroseismology.
In addition, the value of $\beta \simeq 0.25$ we obtained may be too large
for a star with a convective envelope.
These facts, as well as the subtleness of the detected anomaly,
may suggest some incompleteness in the current modeling or 
other origins for the anomaly,
and should be addressed in future studies.

\section{Conclusion}
Our present analysis reproduces the results by B11 with more data and thus
strengthens the reliability of the gravity-darkening method for the spin--orbit angle determination.
In contrast, we also find that the spin--orbit angle obtained from the gravity-darkened transit light curve strongly depends on the assumed limb-darkening profile.
Depending on its choice, the resulting spin--orbit angle can vary by several tens of degrees.
Thus, the reliable modeling of the limb-darkening effect is crucial for this method.

Nevertheless, if $\lambda$ is constrained from other observations, 
$i_\star$ is well determined along with the limb-darkening parameters.
Hence the gravity-darkening method still provides valuable information on the true spin--orbit angle $\psi$,
which is complementary to $\lambda$ from the RM effect or Doppler tomography.
Indeed, such an example is already seen in an eclipsing binary system DI Her \citep{2013ApJ...768..112P}.
In addition, synergy with asteroseismology is also promising
because it constraints $f_{\rm rot}$ and $i_\star$,
which are both essential in the modeling of gravity darkening.
The joint analyses of these kinds may in turn help us 
to better understand the mechanisms of gravity darkening itself,
since they enable the measurements of $\beta$ for stars
not in close binary systems and hence free from the strong tidal distortion.

If combined with continuous, high-precision photometry as achievable with space-borne instruments,
the gravity-darkening method also provides a way to monitor the angular momentum evolution in the system.
Modeling of the spin--orbit precession allows us to access the internal structure of the rotating star through 
its quadrupole moment or moment of inertia.
It is also possible to determine the three-dimensional configuration of the system from a dynamical point of view
\citep[c.f.,][]{2013ApJ...768..112P, 2013ApJ...774...53B}.
Such information will be valuable in simulating the dynamical histories of
individual systems to decipher the origin of the spin--orbit misalignment.

\acknowledgements
The author is grateful to the entire {\it Kepler} team for the revolutionary data.
The author wishes to thank the referee, Jason Barnes,
for his valuable suggestions that improved the manuscript.
The author also thanks Othman Benomar, 
Shoya Kamiaka, Hajime Kawahara, Masamune Oguri, and Yasushi Suto
for helpful discussions and insightful comments.
The data analysis was in part carried out on common use data analysis computer system 
at the Astronomy Data Center, ADC, of the National Astronomical Observatory of Japan. 
K.M. is supported by JSPS Research Fellowships for Young Scientists  (No.\,26-7182) and 
by the Leading Graduate Course for Frontiers of Mathematical Sciences and Physics.



\end{document}